\documentclass[%
 aip,
 jap,%
 amsmath,amssymb,
 reprint,%
]{revtex4-1}

\usepackage{graphicx}
\usepackage{dcolumn}
\usepackage{bm}
\pdfoutput=1

\usepackage{epstopdf}
\usepackage[normalem]{ulem}
\usepackage{xcolor}

\begin{document}

\preprint{AIP/123-QED}

\title{Nanostructural features degrading the performance of superconducting radio frequency niobium cavities revealed by TEM and EELS}

\author{Y. Trenikhina}
 \altaffiliation{yuliatr@fnal.gov}
 \affiliation{ 
Physics Department, Illinois Institute of Technology, Chicago IL 60616, United States
}%

\affiliation{%
Fermi National Accelerator Laboratory, Batavia IL 60510, United States
}%

\author{A. Romanenko}
\altaffiliation{aroman@fnal.gov}
\affiliation{%
Fermi National Accelerator Laboratory, Batavia IL 60510, United States
}%

\author{J. Kwon}

\affiliation{
Materials Science and Engineering Department, University of Illinois, Urbana IL 61801, United States 
}%

\author{J.-M. Zuo}
\affiliation{
Materials Science and Engineering Department, University of Illinois, Urbana IL 61801, United States 
}%

\author{J. F. Zasadzinski}
 \affiliation{ 
Physics Department, Illinois Institute of Technology, Chicago IL 60616, United States
}%


\begin{abstract}

Nanoscale defect structure within the magnetic penetration depth of $\sim$100~nm is key to the performance limitations of niobium superconducting radio frequency (SRF) cavities. Using a unique combination of advanced thermometry during cavity RF measurements, and TEM structural and compositional characterization of the samples extracted from cavity walls, we discover the existence of nanoscale hydrides in electropolished cavities limited by the high field $Q$ slope, and show the decreased hydride formation in the electropolished cavity after 120$^\circ$C baking. Furthermore, we demonstrate that adding $800^\circ$C hydrogen degassing followed by light buffered chemical polishing restores the hydride formation to the pre-120$^\circ$C bake level.
We also show absence of niobium oxides along the grain boundaries and the modifications of the surface oxide upon 120$^\circ$C bake.

\end{abstract}

\maketitle

\section{\label{sec:level1}Introduction}

Superconducting radio frequency (SRF) cavities is the state-of-the-art technology for particle acceleration implemented in most modern and future planned accelerators~\cite{Hasan_book2, Padamsee_Ann_Rev_Nucl_2014}. SRF cavities are predominantly made of bulk niobium and are typically operated at temperatures of 2~K or below, deep in superconducting state of niobium, which has superconducting critical temperature $T_\mathrm{c}=9.25$~K. The performance of SRF cavities is characterized by the maximum accelerating field ($E_\mathrm{acc}$) they can sustain, and the cavity quality factor $Q$ determining their efficiency of operation. Lower $Q$ leads to the increased dynamic heat load for the cryogenic system, and, if severe, can even lead to the limitation in $E_\mathrm{acc}$ as it causes an increase in the inner cavity wall temperature that can trigger the localized loss of superconductivity - quench. The magnitude of $Q$ is determined by the average microwave surface resistance $R_\mathrm{s}$, which consists of the strongly temperature dependent part $R_\mathrm{BCS}(T)$ and a temperature independent (residual) component $R_\mathrm{res}$.

Recent investigations showed that for standard cavity preparation techniques~\cite{Romanenko_Rs_B_APL_2013} as well as for a newly discovered nitrogen doping~\cite{Grassellino_SUST_2013} both $R_\mathrm{BCS}$ and $R_\mathrm{res}$ depend on the surface rf magnetic field magnitude $B \propto E_\mathrm{acc}$. Since these field dependencies are determined by the surface treatments and the magnetic field only penetrates $\sim$100~nm inside niobium in superconducting state at 2~K, the nanostructure within this thickness and its changes with treatments is key to understanding changes in surface resistance and $Q$.

One of the long-standing puzzles is a strong increase in the surface resistance of electropolished cavities above $\approx$100~mT surface magnetic field - a so-called high field $Q$ slope (HFQS). The effect persists in the absence of other well-known parasitic losses such as multipacting and field emission. HFQS can be removed by the empirically found ``mild baking'' at $100$-$120^{\circ}\mathrm{C}$ in ultra high vacuum (UHV) for 24-48 hours~\cite{Hasan_book2}.

Several models for the HFQS were proposed in the past, but most were shown to contradict at least one of the experimental observations~\cite{Ciovati_PRST_Hydrogen}. The most recent promising model is based on the formation of lossy niobium nanohydrides in the penetration depth~\cite{Romanenko_SUST_Proximity_2013}. Nanohydrides may remain superconducting due to the proximity effect up to the breakdown field, which is determined by their size. The model attributes HFQS onset field to such a loss of proximity-induced superconductivity, which manifests as a strong increase in residual resistance and causes HFQS. The rationale for this theory is the presence of high concentration of interstitial hydrogen in the penetration depth~\cite{Antoine_SRF91_Hydrogen, Tajima_SRF03_Hydrogen, Romanenko_SUST_ERD_2011}, which, upon cooling to 2~K, may coalesce into lumps of niobium hydrides. A challenging part is that in order to search for such nanohydrides directly, cryo-investigations at $<$100~K are required as at room temperature no hydrides are present. 

The characteristic feature of the HFQS is the localization of strong additional dissipation in the areas of cavity surface corresponding to highest surface magnetic fields~\cite{Hasan_book2}, as found out by advanced temperature mapping studies of outside cavity walls. A very powerful approach is based on localizing the strongest dissipative spots, which then can be extracted from cavity walls and their inner surface studied by different analytical techniques. Comparing cutouts from the cavities with the HFQS to the ones from the mild baked cavities without HFQS allows drawing conclusions on the possible underlying causes of the HFQS, and the origin of the mild baking effect.

Previous comparative studies on such cavity cutouts provided clues into possible mechanisms at play. Low energy muon spin rotation spectroscopy (LE-$\mu$SR) showed that mild baking leads to a strong decrease in the electron mean free path~\cite{Romanenko_LEM_APL_2014}. Variable energy positron annihilation spectroscopy (VEPAS) showed that inward diffusion of vacancies and the hydrogen-trapping effect of vacancies may be behind this $\ell$ suppression~\cite{Romanenko_VEPAS_APL_2013}. Another alternative mechanism of the mean free path suppression may be based on the inward diffusion of oxygen~\cite{Calatroni_O_Diffusion_SRF_2001}. Both oxygen and vacancies may then serve as trapping centers for hydrogen, preventing the formation of lossy nanohydrides~\cite{Romanenko_SUST_Proximity_2013}.

In this particular work we present structural and analytic comparison of FIB-prepared cross-sectional samples from mild baked and unbaked cavity cutouts using various TEM techniques at room and cryogenic (94~K) temperatures. Temperature dependent nano-area electron diffraction (NED) and scanning electron nano-area diffraction (SEND) reveal the formation of stoichiometric non-superconducting small niobium hydride inclusions. Mild baking is shown to decrease nanohydride sizes and/or density, which directly correlates with the observed suppression of the high field $Q$ slope in SRF cavities. Importantly, we also find nanohydrides after additional 800$^\circ$C degassing for 3 hours followed by buffered chemical polishing for 20~$\mu$m material removal. Furthermore, size and/or density of such hydrides are comparable to the pre-120$^\circ$C bake electropolished cutout. Additionally, high resolution TEM (HRTEM) and bright field (BF) imaging show similar surface oxide thickness and lack of any oxidation along grain boundaries. Electron energy loss spectroscopy (EELS) chemical characterization of the surface oxides suggests slight oxygen enrichment just below the oxides after the mild bake, consistent with the previous X-ray investigations~\cite{Delheusy_APL_2008}.

\section{\label{sec:level1}Methods}

\subsection{\label{sec:level2}Identification of the cutout regions in unbaked and 120$^\circ$C baked cavities}

\begin{figure}[htb]
   \includegraphics*[width=90mm] {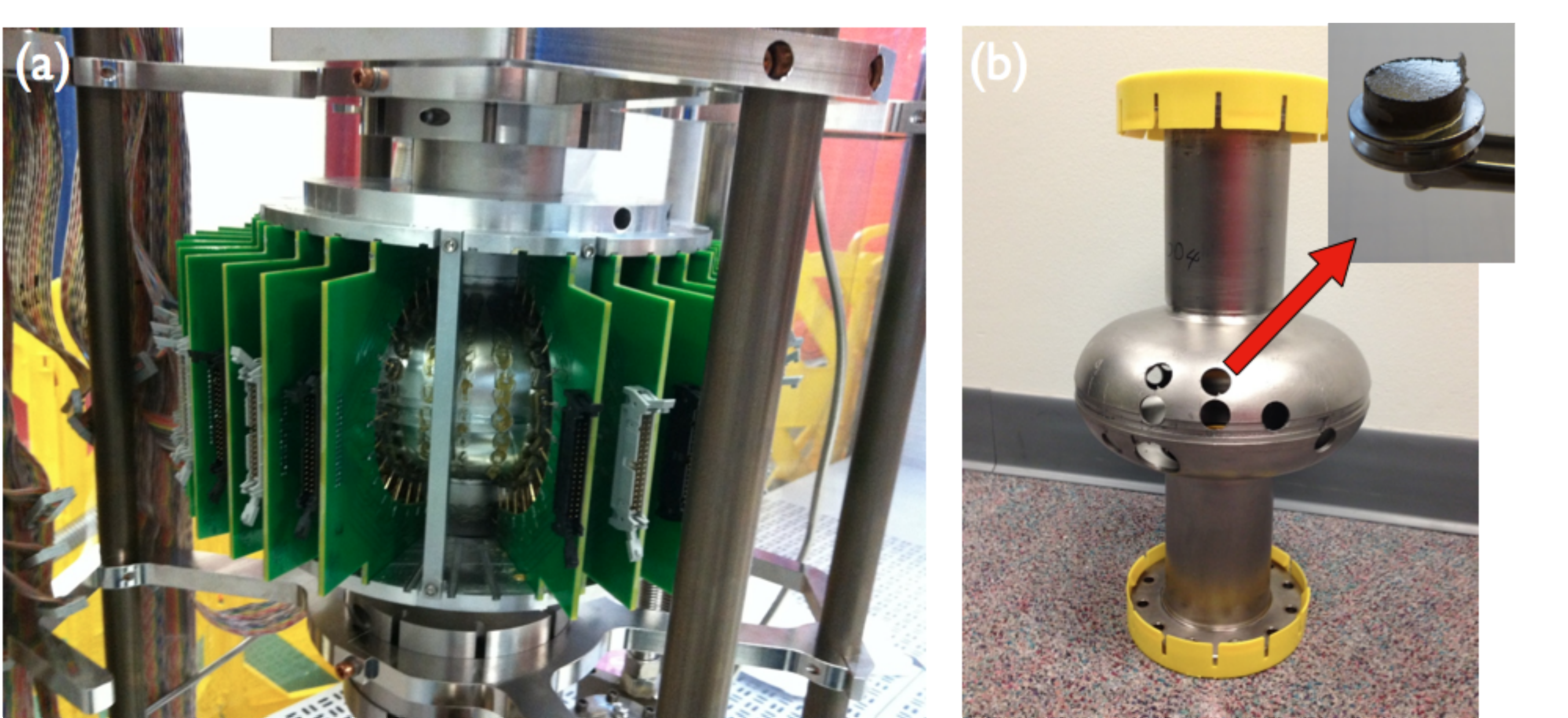}
   \caption{(a) Cavity with attached thermometers. (b) Source cavity with a cutout on SEM post.}
   \label{T-map-pict}
\end{figure}

In order to directly correlate different dissipation characteristics with surface nanostructure and to determine the underlying mechanisms of HFQS in Nb SRF cavities, we base our studies on the comparison of cutouts from cavities with and without HFQS, similar to previous studies~\cite{Romanenko_VEPAS_APL_2013, Romanenko_LEM_APL_2014}. Two Nb fine grain ($\approx$50 $\mu$m) TESLA shape cavities with resonant frequency $f_0=$1.3~GHz were used. Both cavities were electropolished, and one of them was additionally baked at $120^{\circ}\mathrm{C}$ for 48 hours.  Dependence of the quality factor on peak surface magnetic field at 2 K was measured for both cavities, and both also had temperature maps acquired during rf measurements in order to identify the regions for cutout. As expected, the EP-only cavity that had no final mild bake showed prominent HFQS while the performance of the EP+$120^{\circ}\mathrm{C}$ baked cavity was free from the HFQS.

\begin{figure}[htb]
   \includegraphics*[width=90mm] {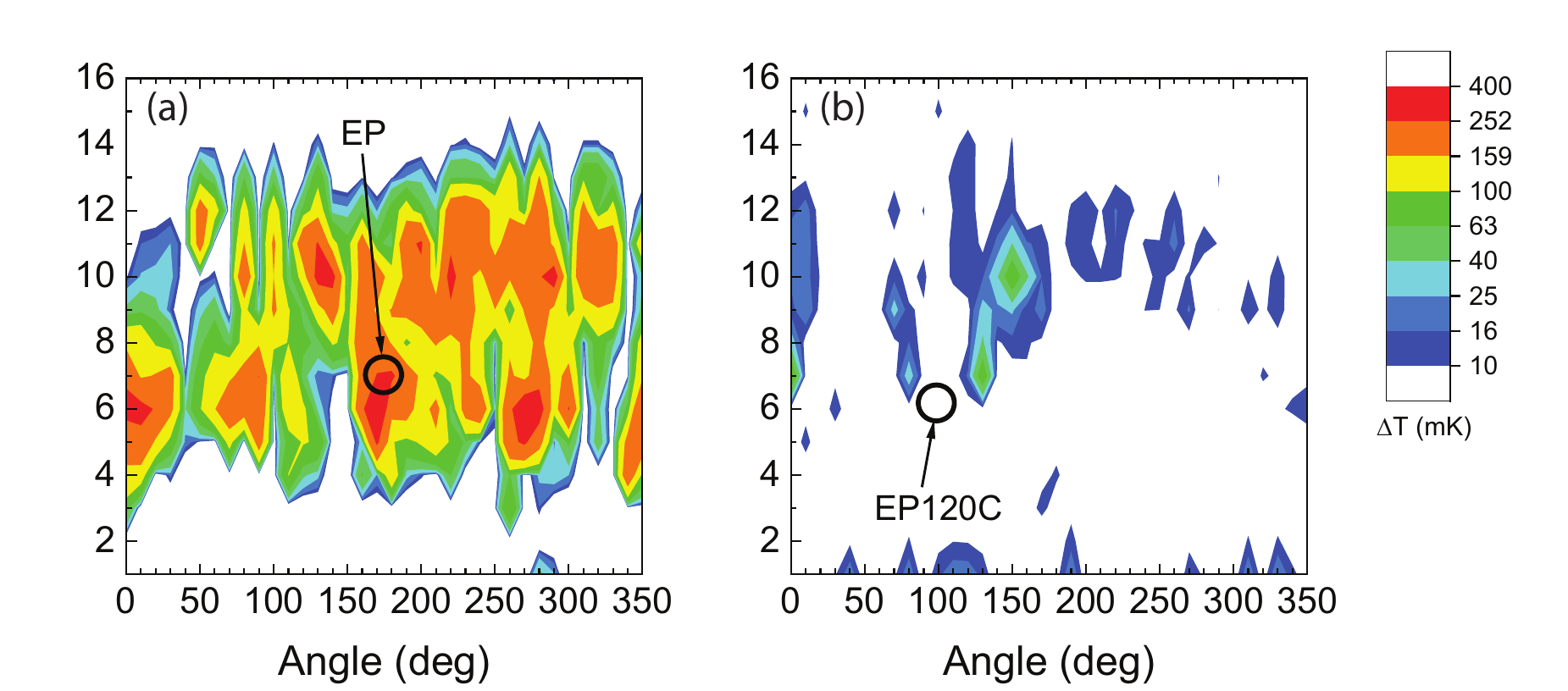}
   \caption{Temperature maps at $E_\mathrm{acc}=28$~MV/m of: (a) unbaked EP cavity, (b) EP+120$^\circ$C baked cavity. Locations of the cutout samples are marked by black circles.}
   \label{T-map}
\end{figure}

\begin{figure}[htb]
   \includegraphics*[width=60mm] {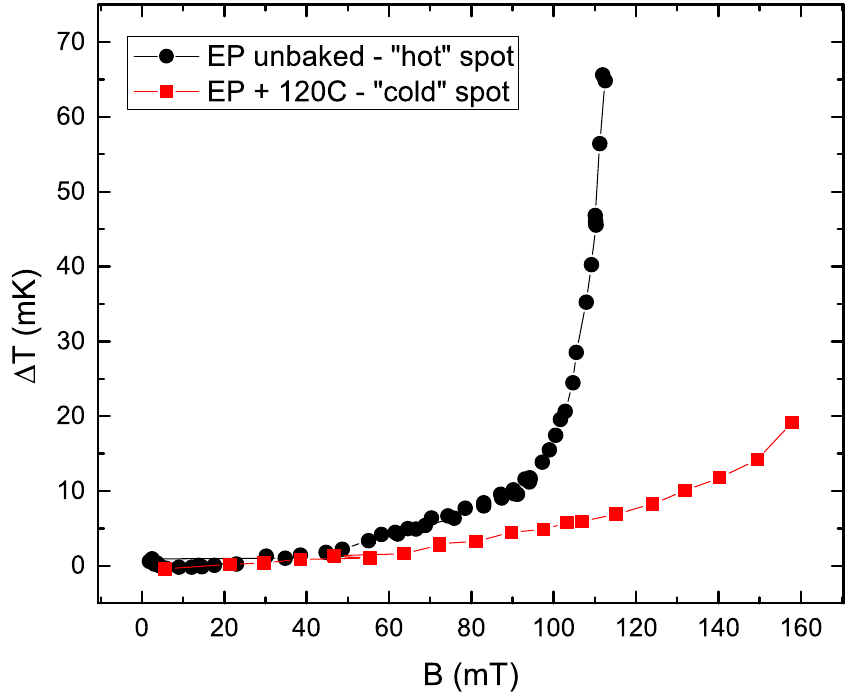}
   \caption{Dissipation characteristics at different rf field amplitudes of EP and EP120C spots.}
   \label{EP-EP120C-Comparison}
\end{figure}

Temperature maps of both cavities recorded at $E_\mathrm{acc}=28$~MV/m, which is above the HFQS onset field, are shown in Fig.~$\ref{T-map}$. The unbaked cavity shows large regions of elevated temperature (up to 0.4 K) - a standard feature of the HFQS - as compared to the cavity that was baked at $120^{\circ}\mathrm{C}$, which had no such extended dissipative regions. Based on the temperature maps, samples with characteristic rf field dependences for each of the treatments (shown in Fig.~$\ref{EP-EP120C-Comparison}$) were cut out from both cavities. The selected characteristic sample from the unbaked cavity (labeled ``EP'' for the following) shows a drastic increase of the local temperature as the surface rf field amplitude reaches about 100~mT. This is in contrast with the characteristic cutout from the $120^{\circ}\mathrm{C}$ baked cavity (labeled ``EP120C'' for the following), which shows no such feature. Cutouts were of circular shape with 11 mm diameter and were extracted from the cavities by an automated milling machine with pure water used as a lubricant.

Additionally, two further samples from the unbaked cavity were subjected to 800$^\circ$C vacuum degassing for 3 hours, and 20~$\mu$m buffered chemical polishing.

\subsection{\label{sec:level2}Characterization of cavity cutouts}

\begin{figure}[htb]
   \includegraphics*[width=90mm] {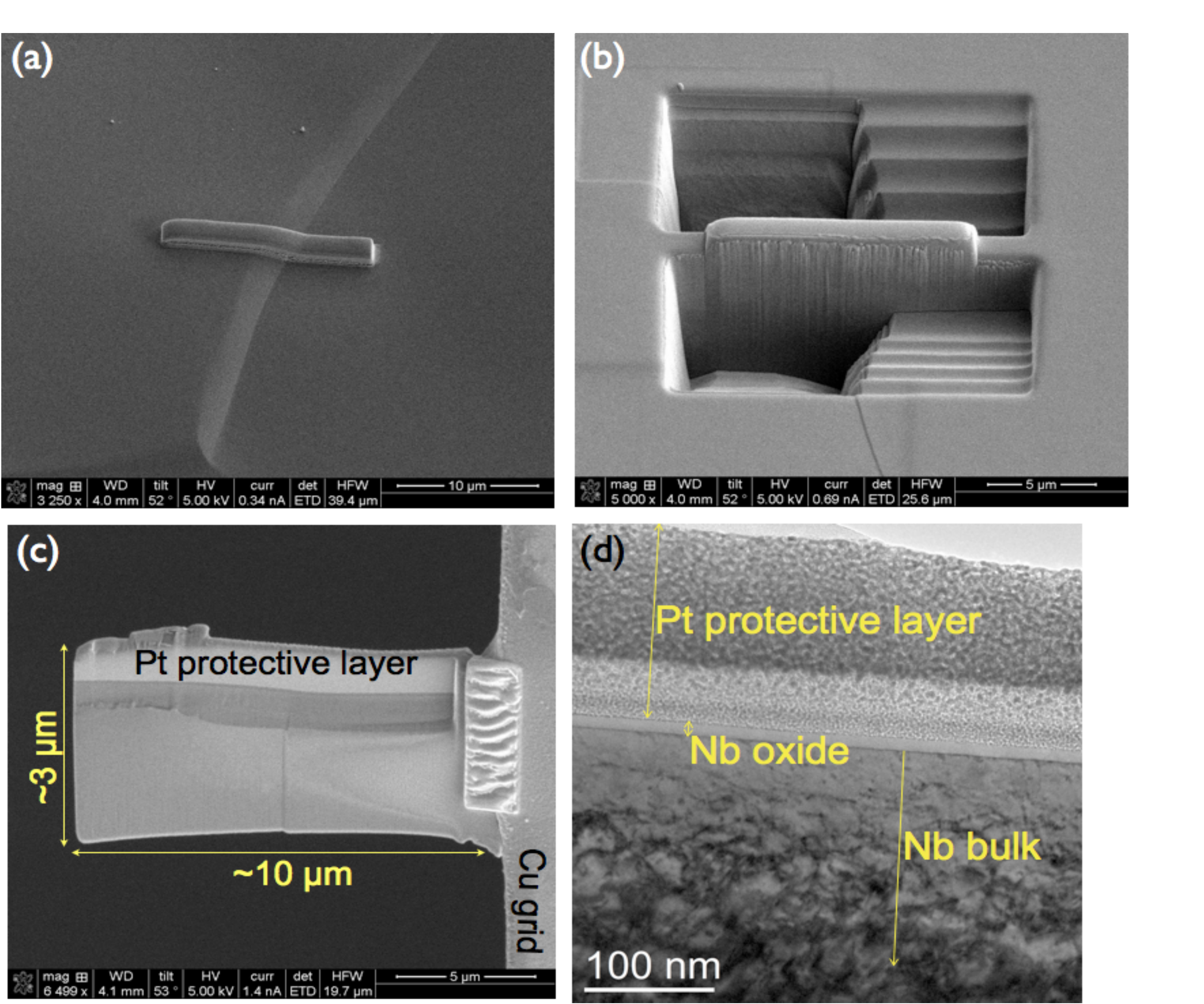}
   \caption{(a) SEM image of protective Pt layer deposition onto Nb surface; (b) SEM image of material removal around cross-sectional rectangular sample; (c) SEM image of TEM FIB-prepared sample on copper grid; (d) TEM image of the sample near-surface.}
   \label{FIB-TEM}
\end{figure}

Cross sectional TEM samples were prepared from cutout samples by Focused Ion Beam (FIB) using a Helios 600 FEI instrument. Conventional electrochemocal polishing methods are not acceptable for Nb cavity investigations because polishing loads the sample with hydrogen.  The FIB lift-out technique allows one to prepare and mount a small rectangular cross sectional sample onto a standard copper TEM half-grid using an Omniprobe micromanipulator.  Before FIB milling, the top surface of each cross sectional cut was covered by a protective layer of platinum, in order to protect the native niobium surface from Ga ions (Fig.~$\ref{FIB-TEM}$).  Most FIB samples were intentionally prepared as Nb bi-crystals in order to overcome possible tilting limitations of the TEM holder.  For every temperature dependent electron diffraction experiment, a new sample was prepared. This precaution was made to avoid possible niobium hydride nucleation centers that could be artificially produced during the fast warm up inside the TEM.

Two types of TEMs were used for this work: field-emission gun (FEG) TEM and thermionic LaB$_6$ gun TEM.  The key difference between these two microscopes is the higher brightness of FEG relative to the thermionic LaB$_6$ gun.  Brightness, which defines the electron dose in a small probe, is a crucial parameter for the investigation of dose-sensitive niobium nanohydrides, as will be described below.

The JEM 2010F Schottky FEG TEM at Materials Research Laboratory (MRL) at the University of Illinois at Urbana-Champaign (UIUC) operated at 197kV and equipped with a Gatan imaging filter (GIF), was used for the temperature dependent NED, SAED, and room temperature EELS.  An approximately 80~nm sized parallel beam was used to record NED patterns onto the Fuji imaging plates.  EELS was collected in Scanning Transmission Electron Microscopy (STEM) mode.  Energy dispersion was set to 0.3 eV/pixel for the niobium M-edge core-loss EELS spectra.  The EELS collection time was set to 10-12 s for each of the spectra.

JEOL JEM 2100 LaB$_6$ thermionic gun TEM at MRL/UIUC was used for temperature dependent NED and scanning electron nano-area diffraction (SEND) \cite{Kim2013, Pennycook}.  In SEND, approximately 170 nm and 100 nm beam sizes were used to obtain diffraction pattern ``maps" by automated rastering of a parallel beam across the area of interest on a FIB sample.  A SEND experiment produces a map of the diffraction patterns which represent a phase in a particular area of the sample.  Automated positioning of the beam was achieved by using a custom-made Digital Micrograph script.  Diffraction patterns were taken sequentially from a specific area of the sample which was imaged prior to scanning.  Diffraction patterns were recorded using a Gatan Ultrascan CCD camera, which is optimized for low contrast biological applications.  The step length of the scans was set equal to the probe diameter to avoid oversampling and gaps in diffraction maps.

JEOL JEM-2100 FasTEM at Northwestern University operated at 200 keV and equipped with GIF was used for room temperature SAED and high resolution TEM (HRTEM) imaging. 

Gatan liquid nitrogen cooled double-tilt stage was used for the low temperature measurements.  FIB-prepared cavity cutout samples were cooled to 94~K inside the TEM in approximately 30 min. Additional 30 min was allowed before the measurements for temperature stabilization. SRF cavities have operational temperatures in the range of 1.2-4.2~K while all hydride precipitation happens during the cool-down at much higher temperatures of $\sim$100-150~K as follows from NbH phase diagram~\cite{Welter_1977} and confirmed by recent direct optical investigations~\cite{Barkov_JAP_Hydrides_2013}. Thus observing at 94~K is fully representative of the hydride state inside niobium at lower temperatures.

\section{\label{sec:level1}results and discussion}

\subsection{\label{sec:level2}Temperature-dependent structural investigations}

\subsubsection{\label{sec:level3}Room temperature measurements}

\begin{figure}[htb]
   \includegraphics*[width=90mm] {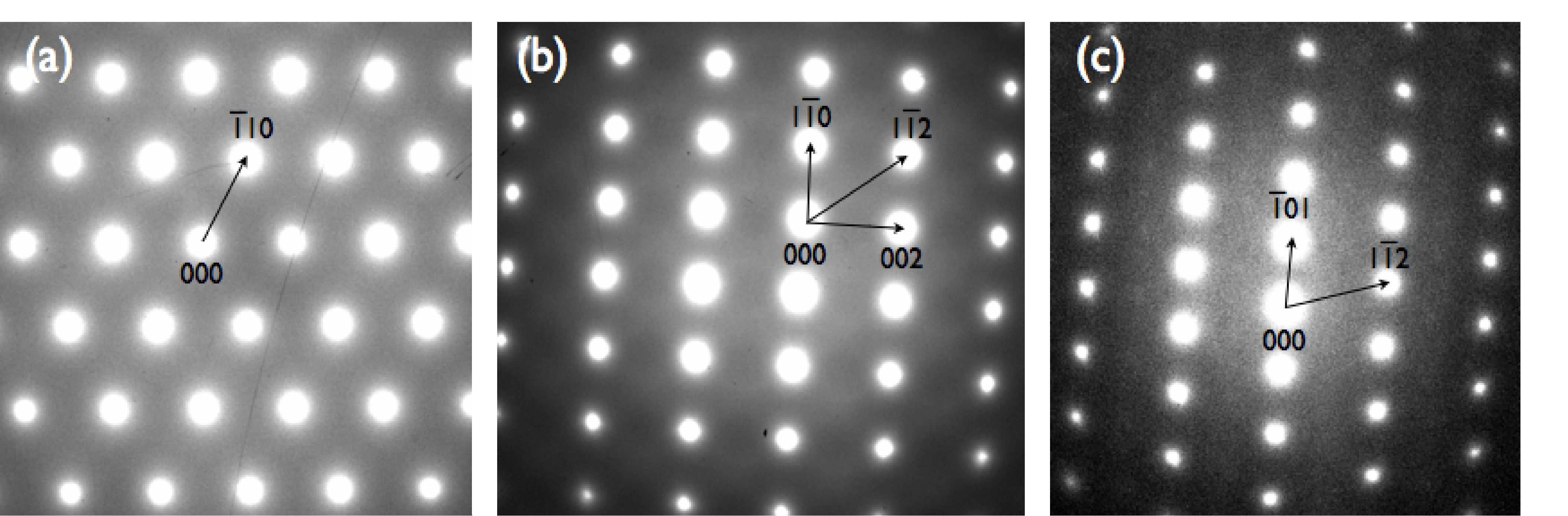}
   \caption{Room temperature NED for EP and EP120C cutouts: (a) Nb [111] zone axis, (b) Nb [110] zone axis, (c) Nb [131] zone axis. }
   \label{NED-roomT}
\end{figure}

Room temperature NED and SAED patterns were first acquired on all of the samples in order to investigate the state of the Nb-H system in the warm state.  NED patterns were taken with a probe size of approximately 80 nm in diameter and areas directly underneath the niobium oxides, as well as areas a few hundred nanometers deep were explored. Similar diffraction patterns produced by body centered cubic (BCC) niobium with no additional ordered stoichiometric phases as shown in Fig.~$\ref{NED-roomT}$ were found on all the samples we investigated.  SAED (not shown), which represents structural information from sample areas of a few micrometers, shows only BCC Nb reflections as well. Thus TEM electron diffraction shows that hydrogen behaves like a lattice gas and occupies random tetrahedral interstitial sites in Nb at room temperature.  This phase is called solid solution ($\alpha$-phase).

\subsubsection{\label{sec:level3}Cryogenic temperature measurements}

Before TEM measurements, in order to confirm the absence of large bulk concentrations of hydrogen, cutouts from the same cavities were investigated in the optical cryogenic stage of the confocal microscope using the same methodology as in our previous study of $Q$ disease-causing larger hydrides~\cite{Barkov_JAP_Hydrides_2013}. No hydride formation was seen on any of the cutouts with the spatial resolution down to $\sim$1~$\mu$m. 

\begin{figure}[htb]
   \centering
   \includegraphics*[width=90mm] {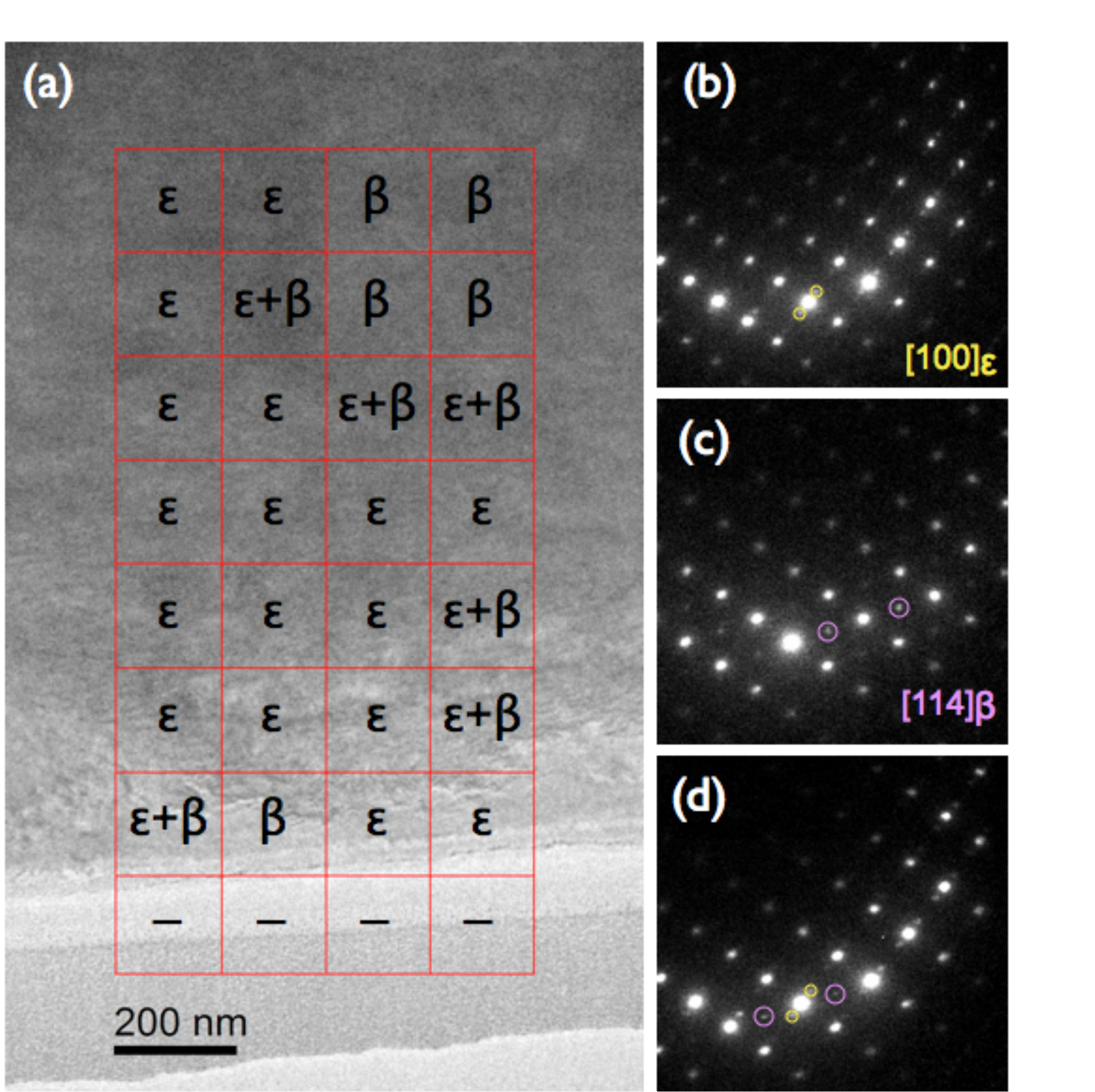}
   \caption{ (a) SEND map of EP sample at 94 K using LaB$_6$ TEM, (b) $\epsilon$-phase Nb$_4$H$_3$ overlapped with Nb, (c) $\beta$-phase NbH overlapped with Nb, (d) $\epsilon$- and $\beta$-phases overlapped with Nb.}
   \label{SEND-EP}
\end{figure}

Cryogenic temperature phase characterization of EP and EP120C samples was accomplished with SEND in thermionic gun TEM and with NED, SAED in FEG TEM.  Fig.~$\ref{SEND-EP}$ shows a SEND map taken from the EP sample along with a TEM image of the sample at 94K.  NED patterns were taken automatically in a sequential manner from the Nb near-surface region which was imaged prior to scanning.  Every square in Fig.~$\ref{SEND-EP}$a represents a sample area of diameter equal to the diameter of the diffraction probe.

\begin{figure}[htb]
   \centering
   \includegraphics* [width=90mm] {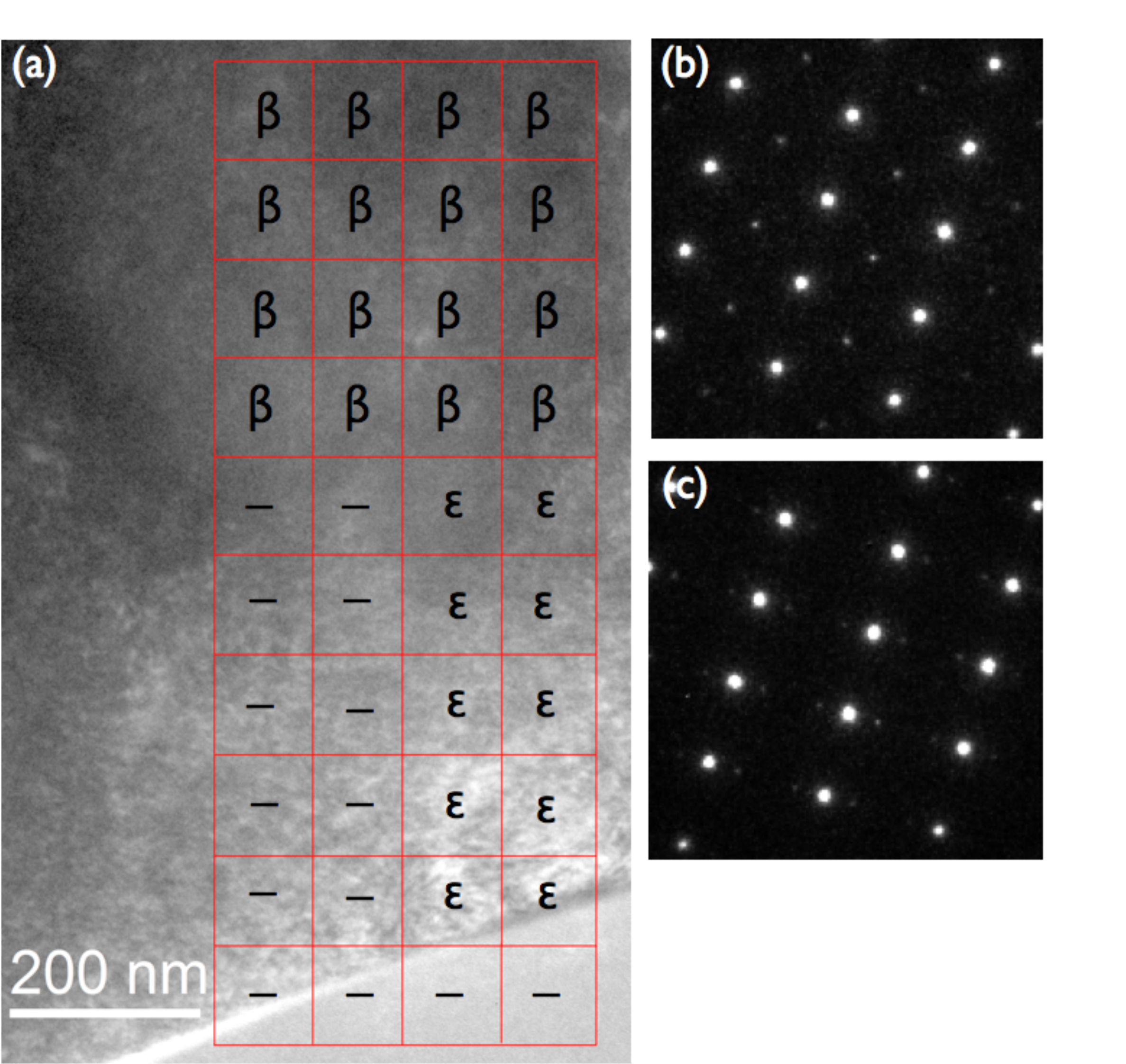}
   \caption{(a) SEND map of EP sample which received 800$^\circ$C vacuum degassing for 3 hours and 20~$\mu$m BCP, measured at 94 K, (b) $\beta$-phase Nb$_4$H$_3$ overlapped with Nb, (c) $\epsilon$-phase NbH overlapped with Nb.}
   \label{800C-SEND}
\end{figure}

Fig.~$\ref{SEND-EP}$b-d show NED representative patterns taken from the ``EP'' sample.  ``Half-order'' additional reflections are clearly visible along with reflections from Nb matrix.  Orientation of Nb crystal is close to [110] zone axis.  Two niobium hydride phases were found in EP sample at 94K.

Fig.~$\ref{SEND-EP}$b shows $\epsilon$-phase niobium hydride diffraction pattern overlapped with [110] Nb.  $\epsilon$-phase was recognized by ``half-order'' reflections along the [110]$_{cubic}$ direction \cite{Schober-lowT}.  $\epsilon$-phase (Nb$_4$H$_3$) has non-centrosymmetric orthorhombic structure with $P_{2_{1}2_{1}2_{1}}$ space group \cite{Schober-lowT, Hauer1998, Somenkov1980}.  The $\epsilon$-phase  forms in $\alpha$ + $\beta$ alloys, by ordering of $\beta$-phase, when H/Nb $<$ 0.7 at 207K \cite{Makenas1982}.  The orientation of observed $\epsilon$-phase domains is close to the [114] Nb$_4$H$_3$ zone axis.  

Fig.~$\ref{SEND-EP}$c shows a $\beta$-phase niobium hydride diffraction pattern overlapped with [110] Nb.  $\beta$-phase was recognized by reflections at $(\frac{1}{2}\ \frac{1}{2}\ 1)_\text{cubic}$ in terms of cubic BCC Nb reflections \cite{Schober-beta}.  $\beta$-phase forms by ordering of the hydrogen interstitials on tetrahedral sites which lie on alternate (1$\bar 1$2)$_c$ planes upon cooling over the composition range of 0.75 $\leq$ H/Nb $\leq$ 1.0 \cite{Makenas1982}.  $\beta$-phase (NbH) has face centered orthorhombic crystal structure with $P_{nnn}$ space group \cite{Somenkov-beta, Schober-beta}.  The orientation of $\beta$-phase domains is close to the [100] NbH zone axis.  

SEND mapping  of EP120C sample at 94K demonstrated only BCC Nb diffraction pattern with no additional reflections.  One EP sample and one EP120C sample were evaluated by SEND at 94K.

\begin{figure*}
   \centering
   \includegraphics[width=160mm] {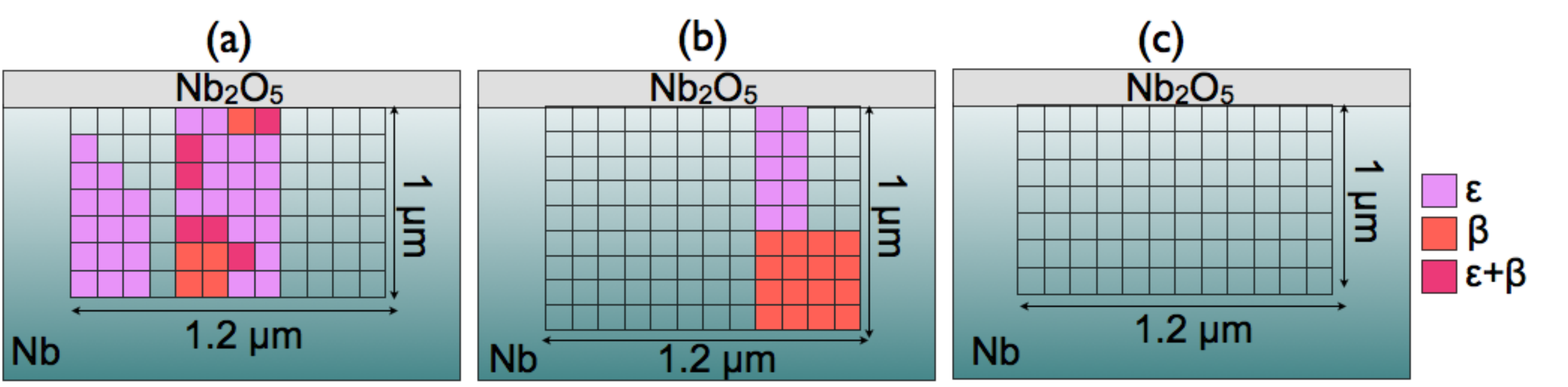}
   \caption{ Diagrams of SEND maps taken at 94K for: (a) EP sample, (b) EP sample which received 800$^\circ$C vacuum degassing for 3 hours + 20~$\mu$m BCP, (c) EP120C sample.}
   \label{diagram}
\end{figure*}

One unbaked cavity sample which received 800$^\circ$C vacuum degassing for 3 hours, and 20~$\mu$m buffered chemical polishing (BCP), was evaluated by SEND at 94K.  SEND mapping (Fig.~$\ref{800C-SEND}$) shows the presence of the same low temperature niobium hydride phases as for the unbaked cavity sample without 800$^\circ$C degassing followed by BCP.  The diagram in Fig.~$\ref{diagram}$ summarizes SEND mapping at 94K for all three types of samples: EP, EP which received 800$^\circ$C degassing and BCP, and EP120C.

\begin{figure}[htb]
   \includegraphics*[width=70mm] {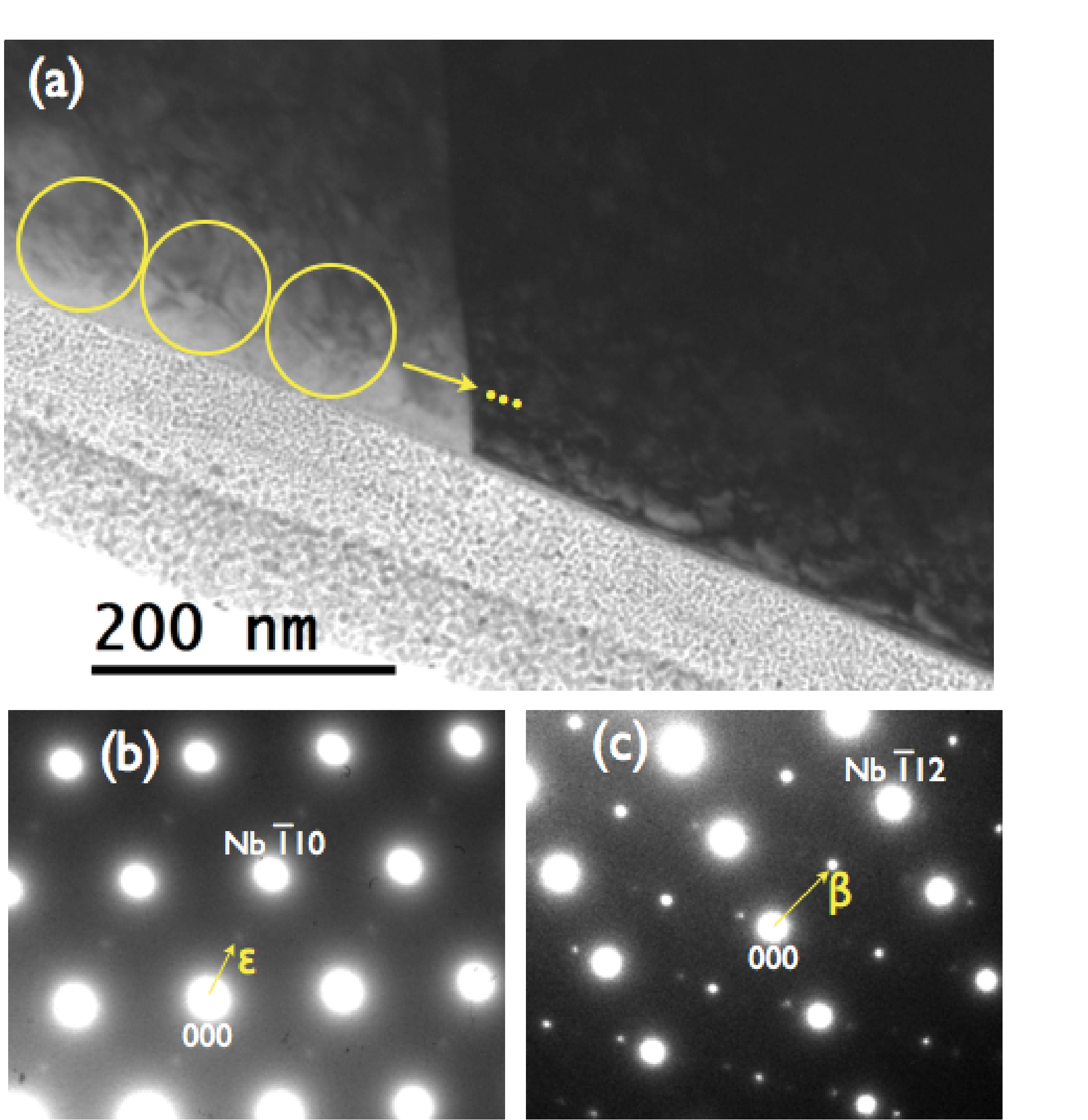}
   \caption{ (a) Bright field image of EP sample.  Right grain is tilted to [110] zone axis.  Positioning of NED probe on TEM sample is indicated by the yellow circles; (b) EP sample NED patterns taken at 94K.}
   \label{TEM-NED-cold}
\end{figure}

\begin{figure}[htb]
   \centering
   \includegraphics[width=70mm] {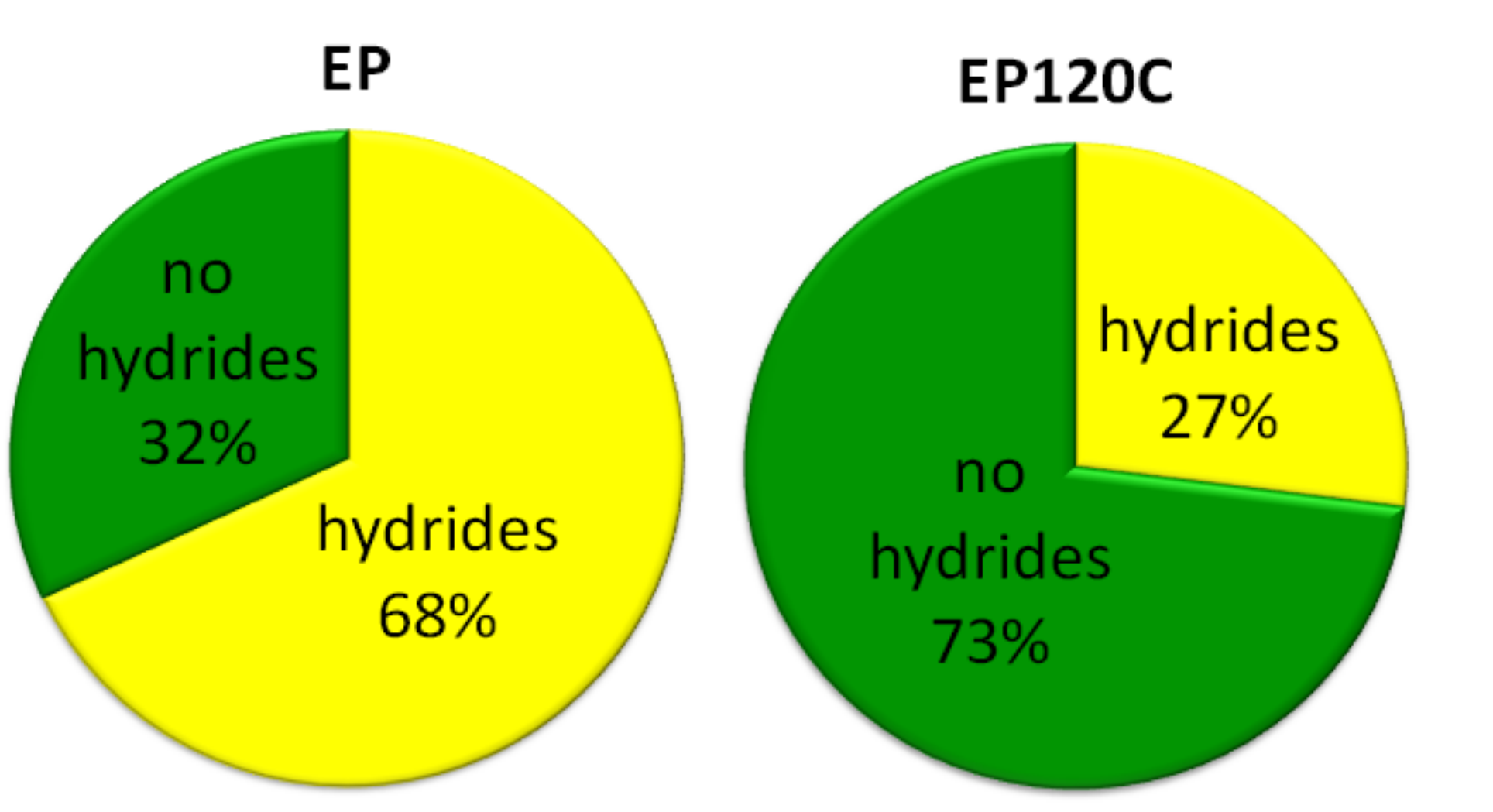}
   \caption{Fraction of area with and without NED-detected hydrides in EP and EP120C samples.}
   \label{NED-pie-chart}
\end{figure}

\begin{figure*}
   \centering
   \includegraphics[width=120mm] {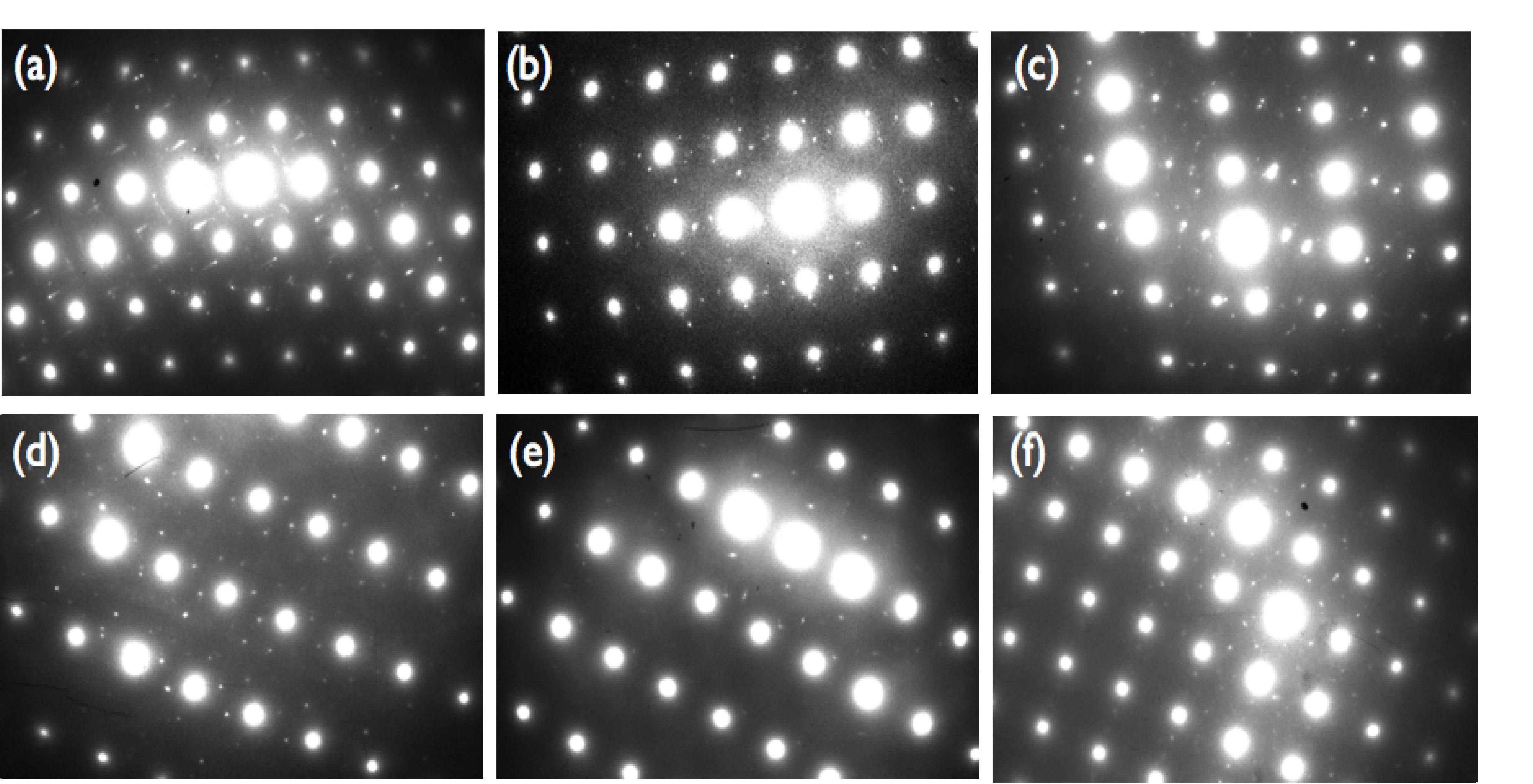}
   \caption{ (a)-(c) Diffraction patterns taken from EP samples at 94K; (d)-(f) Diffraction patterns taken from EP120C sample at 94K.}
   \label{NED-coldT}
\end{figure*}

Additional NED structural characterization of the cavity cutouts was performed in FEG TEM at 94 K.  In order to collect diffraction patterns from the near-surface area, the NED probe was positioned by the deflection coils onto the TEM sample for each exposure.  The size of the NED probe was approximately 80 nm.  NED diffraction patterns were collected by sequentially moving the probe along the length of the sample, which is schematically represented in Fig.~\ref{TEM-NED-cold}a.  Fig.~\ref{NED-coldT} (a)-(c) and (d)-(f) show typical cryogenic temperatures NED patterns taken from EP and EP120C samples, respectively.  Additional second phase reflections are clearly observed along with Nb matrix reflections at 94 K in both types of samples.  Additional low temperature reflections in EP samples are more intense and frequent than additional reflections in EP120C samples.  Comparing EP and EP120C samples, the number of probed spots exhibiting reflections of an additional low-temperature phase differ as shown in Fig.~\ref{NED-pie-chart}.  For the EP samples, 68\% of probed spots showed additional reflections, whereas for the EP120C samples,  27\% of the probed spots showed additional reflections.  Three FIB prepared EP and three EP120C samples were investigated with NED.

Most of the detected second phase reflections were not in compliance with any reported phase of niobium hydride.  Only a few diffraction patterns taken from the EP samples show clear ``half-order" reflections which can be associated with $\beta$- and $\epsilon$-phases of niobium hydrides (Fig.~\ref{TEM-NED-cold}b,c). This can be explained in terms of dissociation of the native low temperature niobium hydride phases under the electron beam exposure.  It has been noticed that additional low-temperature reflections rapidly vanish under exposure to the electron beam.   Heating of the exposed area by electron bombardment or direct electron energy transfer allow hydrogen to regain its mobility and move to different parts of the sample, which can lead to severe distortion and dissociation of niobium hydrides.  We believe that by the time the exposure was taken, the structure of the second phase could be already altered.  A similar effect was previously observed by several researchers~\cite{Schober-lowT, Makenas1982}.  Due to this effect, relevant zone axis tilts were determined prior to cooling, and set up with the minimal sample exposure at low temperatures.

Formation of low temperature stoichiometric niobium hydrides in Nb samples was previously detected by SAED for various hydrogen concentrations~\cite{Tao2013, Schober-lowT, Makenas1982}.  However, low temperature SAED on the samples prepared from the cavity cutouts did not reveal any additional reflections.  The absence of an additional reflections in SAED patterns can be explained either by negligibly small SAED signal from nano-scale niobium hydrides or by their fast dissociation under the broad electron beam.

\subsection{\label{sec:level2}Grain boundaries and surface oxides}

\begin{figure}[htb]
   \centering
   \includegraphics*[width=90mm] {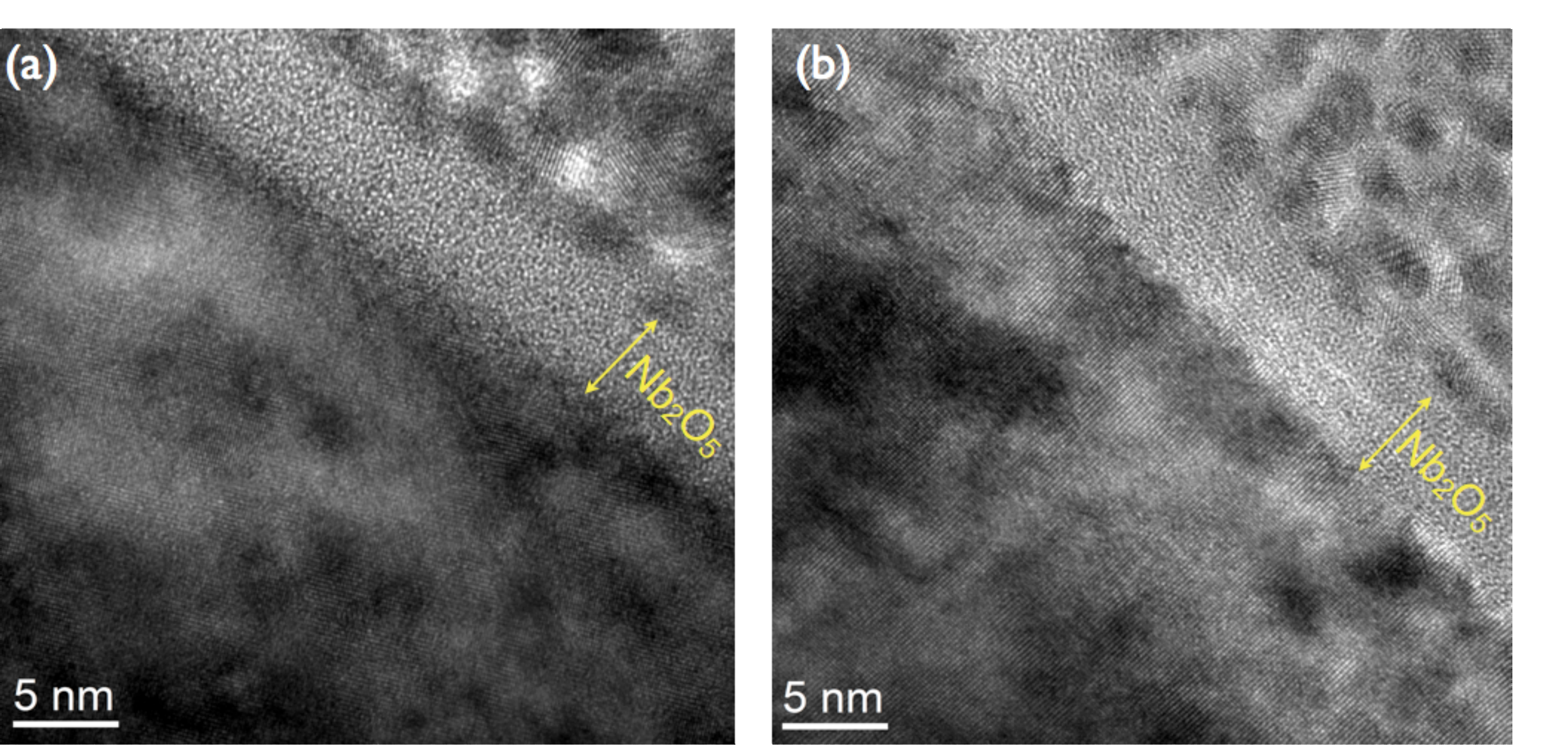}
   \caption{ (a) HRTEM image of EP120C sample under [110] Nb; (b) BF image of EP sample under [110] Nb.}
   \label{HRTEM-2}
\end{figure}

HRTEM imaging and EELS were used for detailed imaging and comparison of the surface oxides and grain boundaries in EP120C and EP samples.  The appearance of an approximately 5~nm-thick amorphous oxide layer (Nb$_2$O$_5$) in HRTEM images is similar for both types of samples (Fig.\ref{HRTEM-2}).

EELS investigations of the surface oxides in the EP120C and EP samples were performed as a way to make detailed comparisons of niobium valence across the oxide layer.   

\begin{figure}[h]
   \centering
   \includegraphics[width=0.5\textwidth] {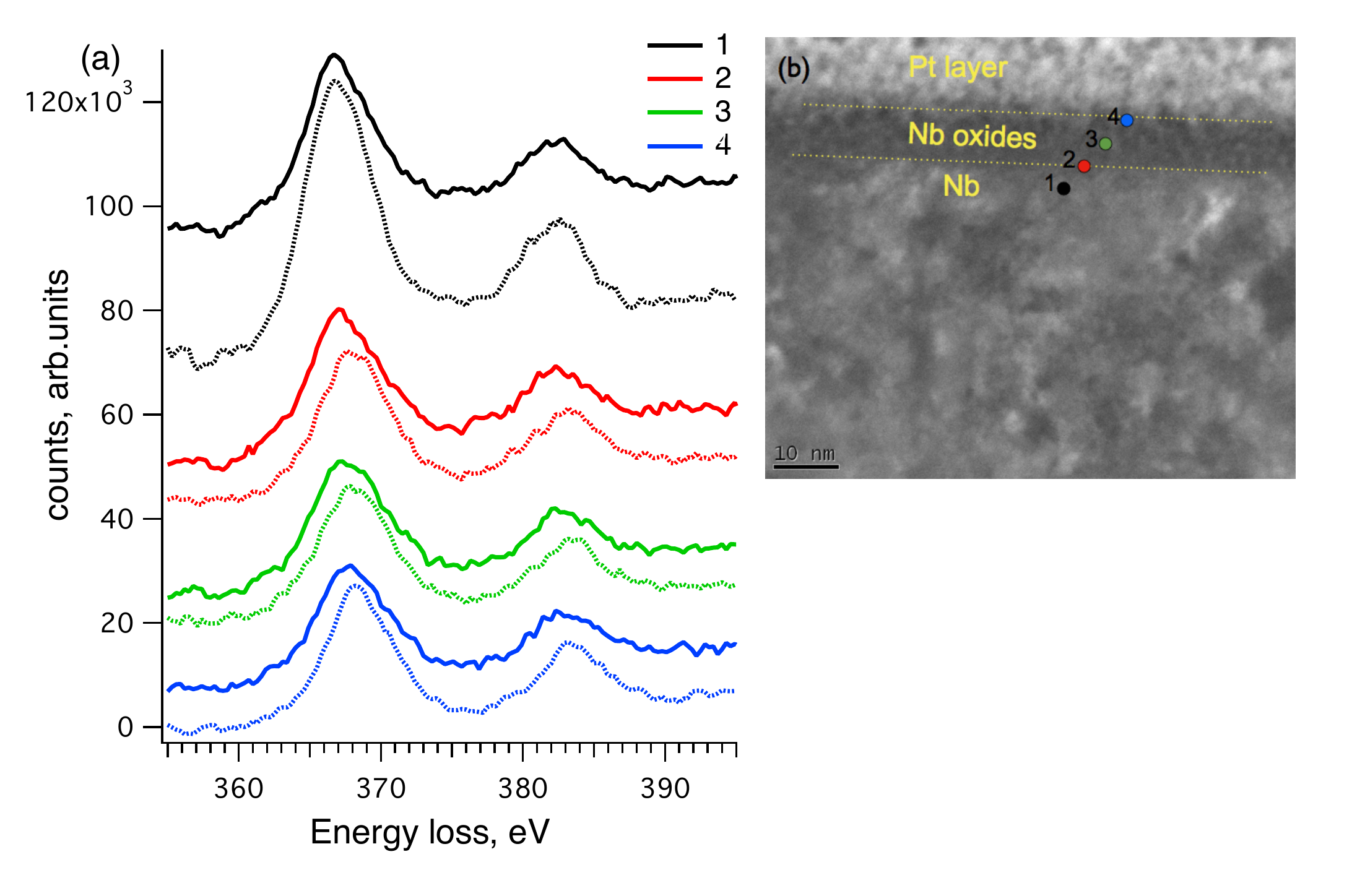}
   \caption{ (a) EELS taken from EP (solid line) and EP120C (dashed line); (b) STEM image indicating the regions where the EELS spectra were taken from.}
   \label{STEM-EELS}
\end{figure}  

Fig.~\ref{STEM-EELS}a shows EELS spectra for niobium M$_{2,3}$ edge.  EELS spectra were taken for the four regions marked in the STEM image of the niobium near surface (Fig.~\ref{STEM-EELS}b).  The M$_{2,3}$ edge of niobium is a result of the transition of Nb 3p electrons to unoccupied Nb 4d and 5s states.  Spin-orbit coupling of the 3p orbital causes the appearance of two peaks (M$_2$ and M$_3$).  All niobium core loss spectra were calibrated with respect to carbon K-edge onset at 286~eV using the second derivative method~\cite{Egerton}.  Three spectra for each region were added after the background subtraction with log-polynomial function~\cite{Bach}.  Thickness of the sample in the region of interest was estimated to be 41~nm from the Log-Ratio Method~\cite{Egerton}.

The linear relationship between the chemical shift of M-edge onset and niobium valence can be used to determine the niobium oxidation state~\cite{Tao2011}.  For both samples, the M$_{2,3}$ peak for each region shows a clear chemical shift toward higher energy as a function of distance from the Nb metallic surface.  The EP120C sample shows a greater shift than the EP sample.  The observed shifts for both the EP and EP120C samples agree well with previous results of Tao et al~\cite{Tao2013}.  Table \ref{M3-positions} summarizes the positions of M$_{3}$ peaks for both samples for each region.  The last column shows the difference in position from region 1 to region 4.  The M$_{2}$ peaks follow the same trend as the M$_{3}$ peaks.  The comparatively larger shifts of the M$_{2,3}$ peaks of the EP120C sample is an indication of higher niobium valence in each region relative to that in the EP sample.  This suggests inward (toward the bulk of Nb) oxygen diffusion during the mild bake.

\begin{table}[ht]
\caption{Experimental position of Nb $M_3$ peak in different regions for EP120C and EP samples.}   
\centering                          
\begin{tabular}{c c c c c c}            
\hline\hline                        
Sample & Region 1 & Region 2 & Region 3 & Region 4 & Shift \\ 
\hline                     
EP  & 366.7 & 367.2 & 367.5 & 367.8 & 1.1\\
EP120C & 366.8 & 368.0 & 367.9 & 368.3 & 1.5 \\
\hline                              
\end{tabular}
\label{M3-positions}                
\end{table}

According to X-ray investigations of Nb/Nb-oxide interfaces~\cite{Delheusy_APL_2008}, an increase of $x$ for NbO$_x$ underneath Nb$_2$O$_5$ in EP120C sample can be caused by the enrichment in interstitial oxygen.  

\begin{figure}[htb]
   \centering
   \includegraphics*[width=90mm] {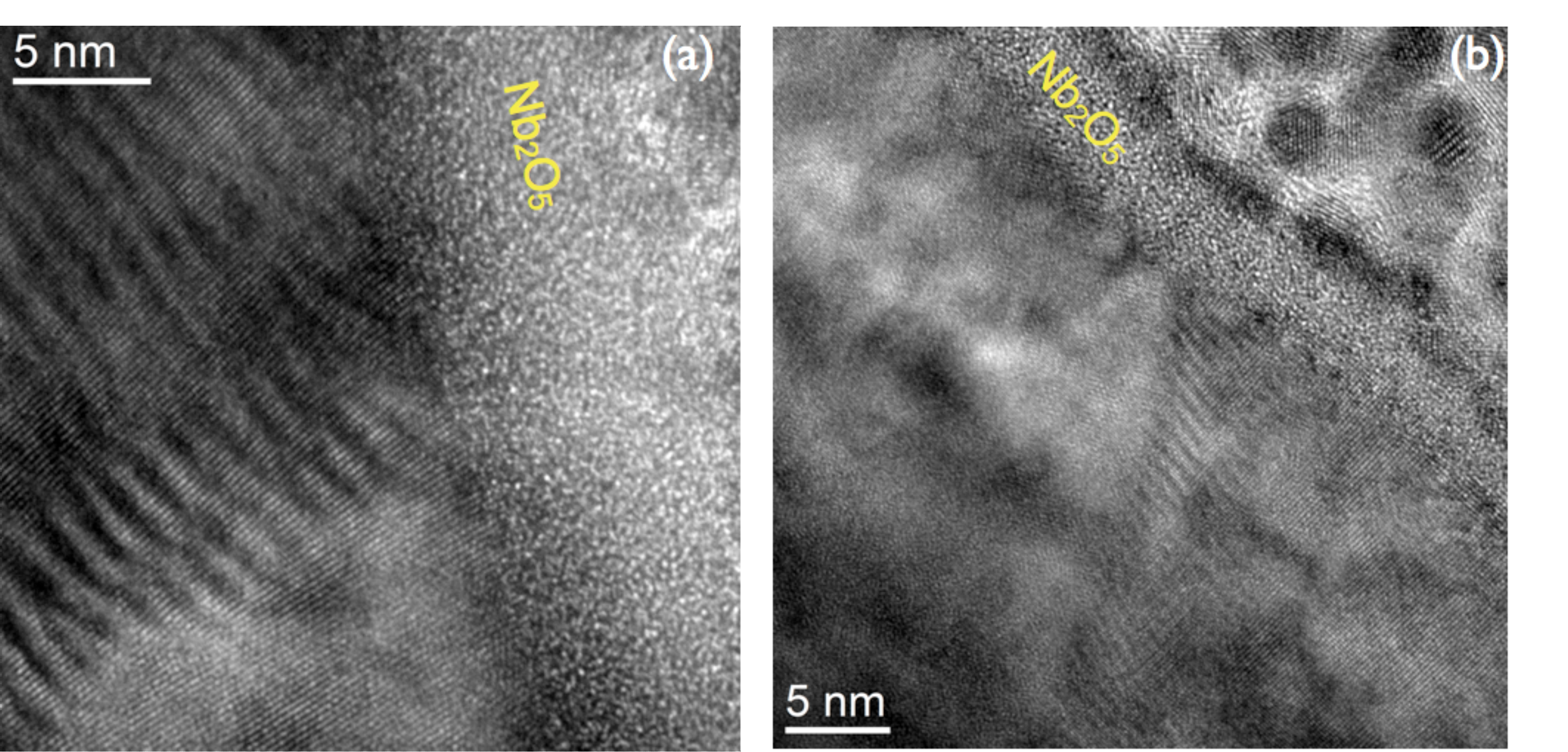}
   \caption{ (a) EP120C sample; (b) EP sample}
   \label{HRTEM-3}
\end{figure}  

Several samples with uniformly thin grain boundary regions were prepared from the cavity cutouts.  Images of the grain boundaries in the EP and EP120C spot samples are represented in Fig.~\ref{HRTEM-3}.  HRTEM images of the cavity cutouts do not show an amorphous contrast from the isolating niobium pentoxide along the grain boundaries, in contradiction to some literature models~\cite{Halbritter1988}.   

\subsection{\label{sec:level2}Comparison of dislocation structure}

\begin{figure}[htb]
   \centering
   \includegraphics*[width=90mm] {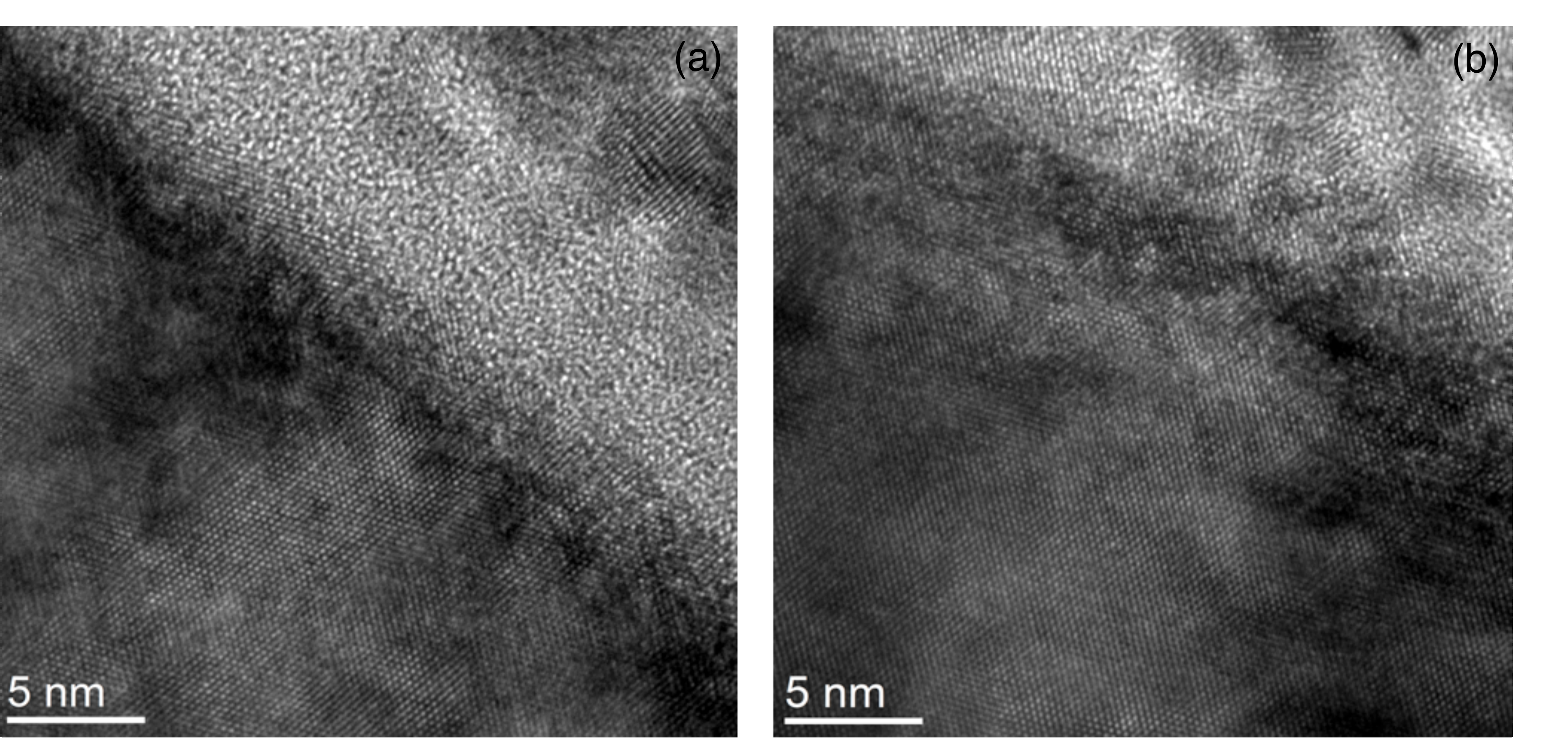}
   \caption{ (a) HRTEM image of EP120C sample; (b) HRTEM image of EP sample.}
   \label{HRTEM-1}
\end{figure}

\begin{figure}[htb]
   \centering
   \includegraphics*[width=90mm] {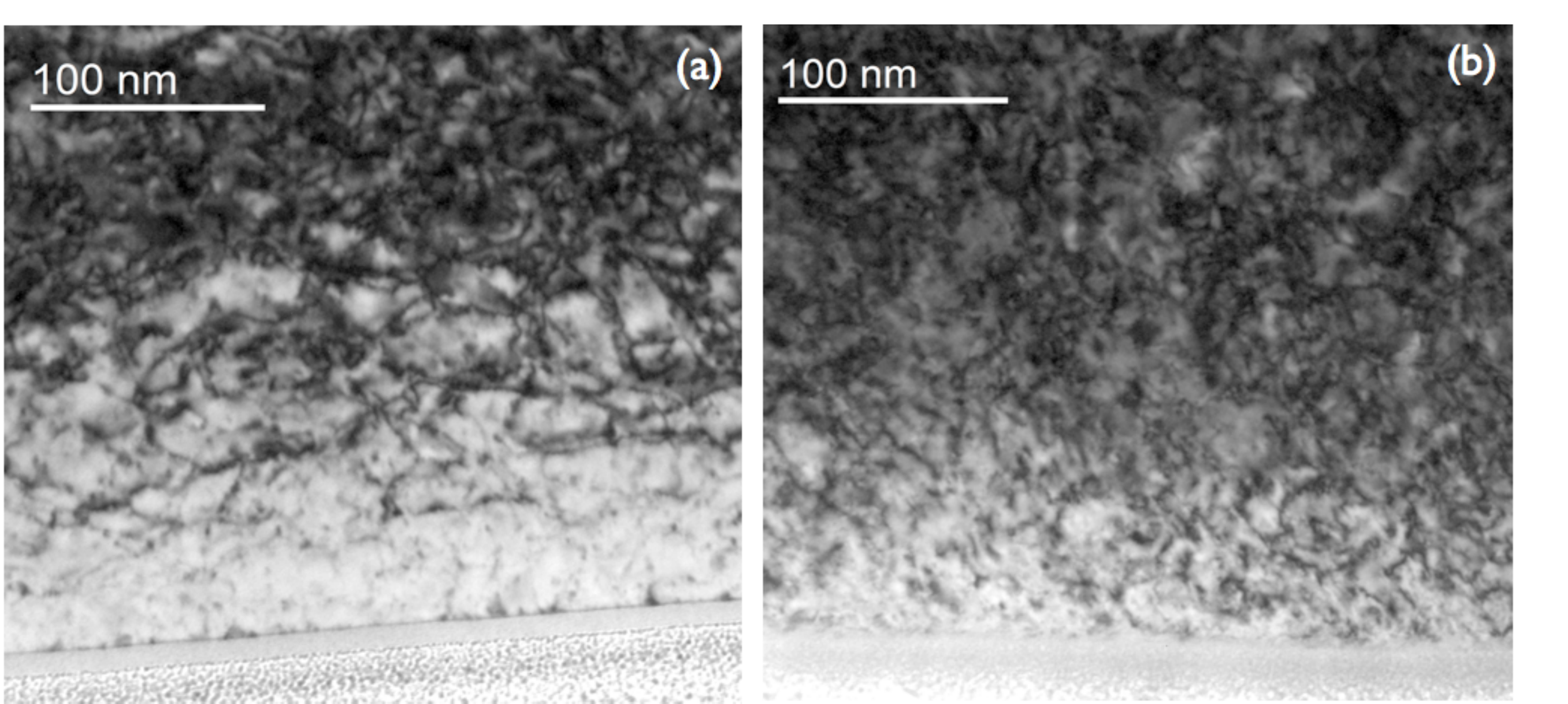}
   \caption{ (a) BF image of EP120C sample under [110] Nb zone axis; (b) BF image of EP sample under [110] Nb zone axis.}
   \label{BF-Hot-Cold}
\end{figure}

Appearance of dislocations produced by the precipitation of niobium hydrides in Nb samples after the first cool down was reported in a number of studies.  Therefore, EP spot samples that suffer from more prominent niobium hydride precipitation at low temperatures can possess higher dislocation density in the near-surface layer at room temperature. To look for this secondary effect, we used HRTEM and Bright Field imaging to compare dislocation content in the EP and EP120C samples.  Fig.\ref{HRTEM-1} shows HRTEM images of EP120C and EP samples under [111] Nb zone axis. Diffraction contrast in Bright Field (BF) images of the EP120C and EP samples confirms a large amount of dislocations in both, which appear as dark streaks and spots (Fig.\ref{BF-Hot-Cold}). This large number of pre-existing dislocations is likely a result of extensive plastic deformation of niobium introduced during cavity manufacturing steps (i.e. deep drawing). Such high dislocation density leads to complicated bending effects which make atomic column projections go in and out of focus in HRTEM images, which made it impossible to discern any effect of NbH precipitation.

\section{\label{sec:level3}Discussion}

Presence and possible involvement of nanoscale niobium hydrides in the HFQS mechanism was recently proposed~\cite{Romanenko_SUST_Proximity_2013} but gaining direct evidence of their existence remained a challenging task. One of the primary findings of our work is the clear demonstration in TEM by NEG and SEND that such nanohydrides do in fact exist, and therefore may indeed be the possible cause of the HFQS. In general, such nanohydrides represent a yet unaccounted for extrinsic mechanism of additional rf dissipation in hydrogen $Q$-disease free SRF cavities, and the full range of their effect on the whole $Q(E)$ curve has to be further understood.

The significantly lower area affected by nanohydride precipitation in EP120C cutouts found by the near-surface NED investigations is consistent with the hydride precipitation suppression by the 120$^\circ$C baking, as proposed recently~\cite{Romanenko_SUST_Proximity_2013, Barkov_JAP_Hydrides_2013, Romanenko_VEPAS_APL_2013}. At this stage, it is not yet possible to definitively say if it is the volume density or the size of the nanohydrides, which is affected. However, the removal of the HFQS suggests that it is likely the size.

Another key finding is the reapperance of high nanohydride population in the cutout sample after additional 800$^\circ$C vacuum treatment for 3 hours followed by 20~$\mu$m buffered chemical polishing. This is a standard processing sequence to guarantee the absence of the $Q$ disease in SRF cavities, which works by drastically lowering the bulk hydrogen content. However, as our results confirm, there is still enough hydrogen near surface - likely due to hydrogen reabsorption in the furnace during cool down and BCP - to cause the formation of nanohydrides. 

It was discussed in the past that niobium oxide structure may also get modified during the 120$^\circ$C bake in several different ways~\cite{Calatroni_O_Diffusion_SRF_2001, Arfaoui_JAP_2002, Ma_Rosenberg_ApplSurfSci_2003, Delheusy_APL_2008}. Our investigations show that amorphous Nb$_2$O$_5$ of about 5~nm thick is very similar in both EP and EP120C cutouts. The slightly increased niobium oxidation state in EP120C is consistent with the increased oxygen concentration right underneath the oxide, as found before~\cite{Arfaoui_JAP_2002, Delheusy_APL_2008}. This increased oxygen concentration may be a reason of the $\sim$1-2~n$\Omega$ higher residual resistance in 120$^\circ$C baked cavities, which can be restored to the pre-120$^\circ$C bake level by the hydrofluoric acid rinse~\cite{Romanenko_HF_PRST_2013} since the oxygen-reach layer gets converted to the newly grown oxide.

Finally, a very important finding is lack of any oxidation along grain boundaries. This contradicts a model of niobium surface, frequently used up to now, which suggests the presence of oxidized grain boundaries, crack corrosion, and isolated niobium suboxides islands~\cite{Halbritter_SRF_2001}. Our investigations show that none of these features are present in SRF cavities.

\section{\label{sec:level1}Summary and Conclusions}

Extensive microscopic comparison of the original cutouts from SRF cavities with and without HFQS was performed in order to elucidate the underlying cause of the HFQS and the mechanism of its cure. TEM comparison using cryogenic NED and SEND of EP and EP120C cutouts revealed for the first time the formation of the near-surface low temperature nanoscale niobium hydride phases, which area density and/or size directly correlates to the presence or absence of the HFQS. Mild $120^{\circ}\mathrm{C}$bake was demonstrated to reduce the amount of and/or change the distribution of niobium hydrides in the near-surface layer.  Phase identification in SEND demonstrated the presence of $\beta$- and $\epsilon$-niobium hydrides in the EP cutout at 94K.  

Additional HRTEM and BF imaging, as well as EELS characterization of the cavity cutouts, were conducted in order to investigate any possible differences in grain boundaries, surface oxides, and dislocation structure.  HRTEM investigation of grain boundaries showed no niobium pentoxide along the grain boundaries and a similar structure of grain boundaries in EP and EP120C samples. 

Identical thickness of the surface niobium pentoxide was found from HRTEM images of the EP and EP120C cutouts.  EELS chemical characterization of the niobium oxidation state as a function of distance from the surface revealed that EP120C samples have higher chemical shifts for all regions, suggesting inward oxygen diffusion from the oxide into the bulk, which may be related to the hydrofluoric acid rinse beneficial effect on 120$^\circ$C baked cavities.
  
\section{\label{sec:level1}Acknowledgement}

The authors would like to thank all the staff scientists who work at Center for Microanalysis of Materials in Frederick Seitz Material Research Laboratory for technical assistance. This work was partially supported by the United States DOE, Offices of Nuclear and High Energy Physics. Fermilab is operated by Fermi Research Alliance, LLC under Contract No. DE-AC02-07CH11359 with the United States Department of Energy.  This work was carried out in part in the Frederick Seitz Material Research Laboratory Central Research Facilities, University of Illinois.  Jihwan Kwon is supported as part of the Center for Emergent Superconductivity, an Energy Frontier Research Center funded by the US Department of Energy, Office of Science, Office of Basic Energy Sciences, under award number DE-AC0298CH10886. The SEND technique was developed with support of DOE BES DEFG02-01ER45923. 

%


\end{document}